\documentclass[iop]{emulateapj}
\slugcomment{Submitted to ApJ 10/24/2014}
\shorttitle{Late-time near-infrared observations of SN~2005df}
\shortauthors{Diamond et al.}

\usepackage{graphicx}
\usepackage[version=3]{mhchem}
\usepackage{aas_macros}
\usepackage{natbib}
\newcommand{\unit}[1]{\ensuremath{\, \mathrm{#1}}}
\begin{document}

\title{Late-time near-infrared observations of SN~2005df}
\author{Tiara R. Diamond $^1$}
\author{Peter Hoeflich $^1$}
\author{Christopher L. Gerardy$^{1,2}$}
\affil{$^1$Department of Physics, Florida State University}
\affil{$^2$Department of Physics, University of North Carolina, Charlotte}

\begin{abstract}
We present late-time ($200-400\unit{day}$) near-infrared spectral evolution for the Type Ia supernova SN~2005df.
The spectra show numerous strong emission features of [\ion{Co}{2}], [\ion{Co}{3}] and [\ion{Fe}{2}] throughout the 
$0.8-1.8\unit{\mu m}$ region.
As the spectrum ages, the cobalt features fade as would be expected from the decay of \ce{{}^{56}Co} to 
\ce{{}^{56}Fe}. 
We show that the strong and isolated [\ion{Fe}{2}] emission line at $1.644\unit{\mu m}$ provides a unique tool to 
analyze near-infrared spectra of Type Ia supernovae.
Normalization of spectra to this line allows separation of features produced by stable versus unstable isotopes of 
iron group elements. 
We develop a new method of determining the initial central density, $\rho_c$, and the magnetic field, $B$, of the 
white dwarf using the width of the $1.644\unit{\mu m}$ line.
The line width is sensitive because of electron capture in the early stages of burning, which increases as a 
function of density. 
The sensitivity of the line width to $B$ increase with time and the effects of the magnetic field shift towards later 
times with decreasing $\rho_c$.
The initial central density for SN~2005df is measured as $\rho_c=0.9(\pm0.2)$ (in $10^9\unit{g}\unit{cm^{-3}}$),  
which corresponds to a white dwarf close to the Chandrasekhar mass ($\rm M_{Ch}$) with 
$\rm M_{WD}=1.313(\pm0.034)\unit{M_{\sun}}$ and systematic error less than $0.04\unit{M_{\sun}}$.
Within $\rm M_{Ch}$ explosions, however, the central density found for SN~2005df is very low for a 
\ce{H}-accretor, possibly suggesting a helium star companion or a tidally-disrupted white dwarf companion.
As an alternative, we suggest mixing of the central region.
We find some support for high initial magnetic fields of strength $10^6\unit{G}$ for SN~2005df, however, 
$0\unit{G}$ cannot be ruled out because of noise in the spectra combined with low $\rho_c$.
\end{abstract}


\section{Introduction}
\label{sec:intro}

Type Ia supernovae (SNe~Ia) are extremely important in astronomy because of their use in measuring 
cosmological distances and the opportunities they provide to study the physics of flames, instabilities, radiation 
transport, non-equilibrium systems, and nuclear and high energy physics.
SNe~Ia are bright time-dependent events which have been seen out to redshifts of $z=1.71$ \citep{rubin13}.
Although there is some variety in the absolute luminosity of these events, they have been shown to be 
standardizable by using characteristics of their light curves \citep{phillips93}.
With increasing numbers of SNe~Ia being discovered, differences in spectra can be examined in order to create 
additional corrections to the standardization of these objects and to further hone our understanding of the 
progenitor systems, the cause of the explosion, and the propagation of material and energy due to the explosion.
For overviews see \citet{branch95}, \citet{nomoto03}, \citet{stefano11}, \citet{stefano12}, \citet{wang12}, and 
\citet{hoeflich13}.

The consensus is that SNe~Ia come from close binary systems and result from a degenerate carbon/oxygen (CO) 
white dwarf (WD) undergoing a thermonuclear runaway \citep{hoyle60}.
There are multiple potential progenitor systems in addition to multiple explosion scenarios.
In the double degenerate (DD) system, the binary system consists of two or more WDs.
In the single degenerate (SD) system, the binary system may consist of a single WD and either a main sequence, 
helium star, or giant star companion
\citep{nomoto84,hernanz88,piersanti03,loren09,piersanti09,pakmor12,wang12,hoeflich13,tornambe13}.
Whether SNe~Ia occur from one or likely both of the proposed progenitor systems, the rate of SNe~Ia from each 
channel and observations of either the progenitor system or the post-explosion companion in supernova remnants 
(SNR) will help solidify the overall picture of SNe~Ia.
Observations of both SD and DD candidate systems can be found in \citet{greiner91}, \citet{vandenheuvel92}, 
\citet{rappaport94}, \citet{kahabka97}, \citet{ruiz-lapuente04}, \citet{gonzalez09}, \citet{kerzendorf09},  
\citet{edwards12}, and \citet{schaefer12}, and, in particular, \citet{foley14} and \citet{mccully14}.

Aside from the makeup of the progenitor system, the explosion mechanism is also a topic of debate. 
In the case of two WDs merging on a dynamical time-scale, the explosion is  triggered by heat release during the 
merging process \citep{iben84,webbink84,benz90,loren09} (only possible in a DD system). 
Alternatively, the explosion can be triggered by compressional heat in an accreting WD when it approaches the 
Chandrasekhar mass limit ($\rm M_{Ch}$).
The donor star can be a non-degenerate star or a tidally disrupted WD \citep{whelan73,wang12,hoeflich13}.
Scenarios with $\rm M_{Ch}$ explosions, whether originating from SD or DD progenitor systems, seem to be 
favored by observations of light curves and spectra, with some additional contribution from dynamical mergers 
\citep{saio85,hoeflich96,saio98,shen12}.
Uncertainty about the progenitor system and explosion mechanism results in questions about the WD central 
density, total mass attained by the WD prior to the explosion, and the density profile of the WD just prior to the 
explosion.

The telltale indicator between the explosion scenarios is the central density of the WD at the time of the explosion.
Within $\rm M_{Ch}$ models, the central density is an additional free parameter which depends mostly on 
the accretion rate and history
\citep{sugimoto75,nomoto84,thielemann86,brachwitz00,hoeflich06a,hoeflich10,seitenzahl11}.
Slower accretion rates will lead to higher $\rho_c$ because electron conduction is able to remove heat from the 
WD core and delay ignition of \ce{C} in the core.
In dynamical or violent mergers, the time scale for merging does not allow for cooling and ignition can be triggered 
at a much lower $\rho_c$.

Nuclear burning of a CO WD provides the energy to overcome the binding energy and provides the kinetic energy 
of the explosion.
The time scales and the burning products depend mostly on the local density at the time of burning.
Hydrodynamical instabilities may cause a redistribution of the burning products.  
Within the $\rm M_{Ch}$ channel, the most likely scenario involves delayed-detonation models 
\citep{khokhlov91b,yamaoka92,woosley94,gamezo04,roepke07,sim13}, i.e., models with a transition from a 
deflagration front to a detonation front (delayed-detonation transition; DDT).  
Independent of the explosion model, burning of a \ce{C}/\ce{O} mixture will produce regions of: iron group elements 
when burning reaches nuclear statistical equilibrium (NSE); layers of \ce{S}/\ce{Si}/\ce{Ar}/\ce{Ca} by incomplete 
burning; \ce{O}/\ce{Mg}/\ce{Ne} as a product of explosive carbon burning; and some unburned outer layers.
In the inner layers, NSE is achieved in all current explosion scenarios.
However, only in $\rm M_{Ch}$ does burning occur under sufficiently high densities that electron capture can 
take place on hydrodynamical time scales which, in turn, shifts the peak of NSE from \ce{{}^{56}Ni} to stable 
isotopes of the iron group. 

Observations during the photospheric phase allow investigation of the layers of incomplete burning and the 
\ce{{}^{56}Ni} region, which determine the peak brightness and $\Delta m_{15}$.
However, observations of these regions provide hardly any information on the central region. 
The isotropic structure of the central region can be probed during the nebular phase by line profiles. 
To use line profiles we need an emission feature of iron that is unblended. 
Optical and most near-infrared (NIR) features are dominated by blends of multiple bound-bound transitions of various elements.
However, the $1.644\unit{\mu m}$ feature of [\ion{Fe}{2}] is a sensitive tool to separate stable and radioactive 
isotopes \citep{hoeflich04b,motohara06,maeda10,sadler12,penney14}.
To produce an emission line, we need both the element and energy input by radioactive decay. 
After $200-300\unit{days}$, the energy deposition is dominated by positrons, which cannot excite the central 
low-velocity non-radioactive regions. 
The low-velocity emissivity produces ``flat-topped'' or ``stubby'' profiles which are a signature of $\rm M_{Ch}$ 
explosions without mixing. 
Up to about $200\unit{days}$, the line profiles are peaked because gamma-rays contribute significantly to the 
energy input and excite the central, non-radioactive iron. 
Time evolution of the line profiles after $2-3\unit{years}$ is sensitive to positron transport effects and, therefore, 
magnetic fields.
Time evolution of the spectrum is needed in order to separate density, asymmetry in chemistry, and magnetic fields 
effects on the line profiles. 

For the first time, a late-time evolutionary sequence of a SNe~Ia is presented in the NIR that includes the 
$1.644\unit{\mu m}$ line of [\ion{Fe}{2}].   
Details of SN~2005df are presented in Sections~\ref{sec:obs-red} and \ref{sec:pre-analysis}.
Reference models and rationale are outlined in Section~\ref{sec:models}.  
We develop general methods for comparing SNe~Ia with the reference models as well as discuss the interpretation 
of the SN~2005df data in terms of our reference models and the implications of the analysis in 
Section~\ref{sec:results}.
We compare the entire spectra between $0.82-1.8\unit{\mu  m}$ to show the viability and uniqueness of the 
$1.644\unit{\mu m}$ [\ion{Fe}{2}] feature. 
Finally, we discuss the implications of our results in Section~\ref{sec:conclusions}.

\section{Observations of SN~2005df and Data Reduction}
\label{sec:obs-red}

SN~2005df was discovered on August 4, 2005, by \citet{cbet192,iauc8580}.
Mid-infrared (MIR) observations of this SN Ia were previously analyzed by \citet{gerardy07}, focusing on the 
[\ion{Co}{3}] emission line at $11.89\unit{\mu m}$ and the [\ion{Ar}{2}] and [\ion{Ar}{3}] emission lines at 
$6.99\unit{\mu m}$ and $8.99\unit{\mu m}$, respectively.
Reviewing the initial published photometry and spectroscopy of SN~2005df, in addition to the range of distance 
estimates to the host galaxy NGC~1559, that analysis seemed consistent with a somewhat sub-luminous SN~Ia 
\citep{gerardy07}.
Light curve data for SN~2005df in the UV were analyzed by \citet{brown10} and \citet{milne10}.
The supernova's peak date in B-band was August 15, 2005.
\citet{milne10} found that the light curve and $\Delta m_{15}$ value of SN~2005df were consistent with a normal 
bright SN~Ia, with a distance to the host galaxy larger than that used in the \citet{gerardy07} analysis.

For our analysis, observational epochs are given in relation to the estimated explosion date of August
2, 2005, to maintain consistency with \citet{gerardy07}.
Four epochs of late-time near-IR spectroscopy, ranging from $198$ to $380\unit{days}$ after explosion, were
observed  using the Gemini Near Infrared Spectrograph (GNIRS) at the Gemini South Observatory.
All spectra were taken in cross-dispersed (XD) mode, with the last observation also including spectra taken in 
long-slit (LS) mode.
The spectra taken in the XD mode are spread out into different spectral orders; however, only orders $3-8$ have 
significant signal and are used in this analysis.
The data from the $198\unit{day}$ observation were combined with the $199\unit{day}$ observation, because they 
lack accompanying standard star calibration images.
The data were reduced following the standard procedure outlined in the Gemini IRAF package, with an exception 
for variance handling.
Because of the low signal-to-noise ratio of these data, the variance was propagated for each step in the reduction 
using a custom script, assuming the original variance determined by the Gemini IRAF package and Poisson 
distribution of noise.
This variance propagation deviates from the standard procedure during the \textsc{nscombine} and 
\textsc{nsextract} steps in the reduction.

The telluric standard stars we use for calibration purposes are SAO 248743, used with the $198\unit{day}$ and
$199\unit{day}$ observations, and SAO 249308, used with the $217\unit{day}$ and $380\unit{day}$ observations.
Both of these stars have a spectral type of A0, with visual magnitudes of $8.3\unit{mag}$ and $7.3\unit{mag}$, 
respectively \citep{ochsenbein00}.
We use Kurucz' model for an A0V star along with the reduced telluric standard star spectra to determine the 
atmospheric and detector response corrections for the data \citep{kurucz93}.
The resolution for the theoretical standard star are at levels of $20\unit{\AA}$, $50\unit{\AA}$, and $100\unit{\AA}$
in the $<1.0\unit{\mu m}$, $1.0-1.6\unit{\mu m}$, and $>1.6\unit{\mu m}$ regions, respectively.
The $198\unit{day}$ and $199\unit{day}$ observations are corrected to be at the same relative fluxing as the
$217\unit{day}$ and $380\unit{day}$ observations, and then normalized to the $217\unit{day}$ observation's
[\ion{Fe}{2}] peak at $1.644\unit{\mu m}$.
Absolute flux calibrating is not possible because photometric observations were not included in the observations. 
Fluxes have been normalized to the peak of the emission feature at $1.64\unit{\mu m}$ for this analysis.

\section{Preliminary Analysis}
\label{sec:pre-analysis}

\begin{figure}[h]
\centering
\includegraphics[width=0.5\textwidth]{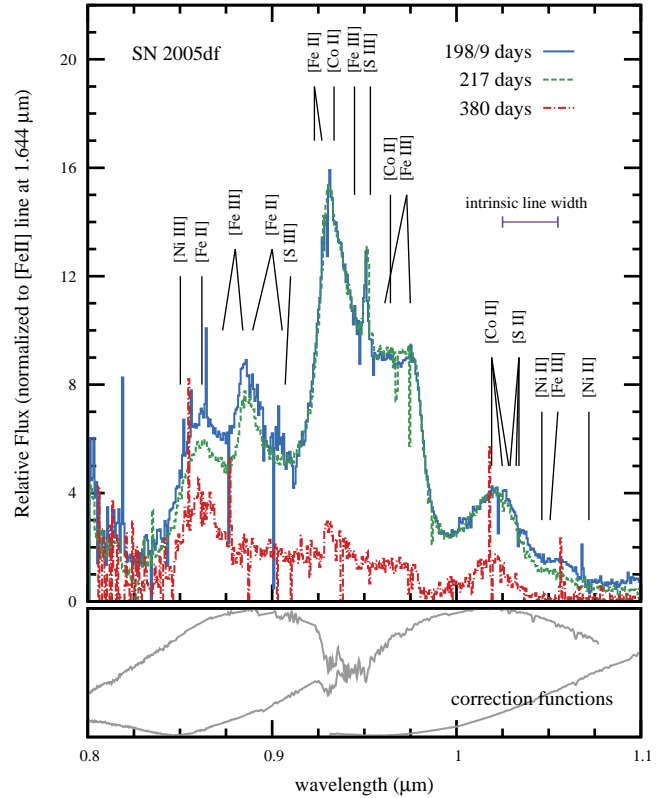}
\caption{Spectra of SN~2005df in the region $0.8-1.1\unit{\mu m}$ at $198/9\unit{days}$, $217\unit{days}$,  and 
$380\unit{days}$.
The spectrum is dominated by forbidden transition emission lines of iron, nickel, cobalt, and sulfur. 
Line identifications and the intrinsic line width are based on spectra from the reference model and its \ce{{}^{56}Ni} 
distribution, respectively, as discussed in Figure~\ref{sec:models}.
Note that all of the features are heavily blended.
In the lower plot,  the relative correction factors are given for the atmosphere and the detector response for orders 8, 
7, and 6.
}
\label{fig:sci_6-8}
\end{figure}

\begin{figure}[h]
\centering
\includegraphics[width=0.5\textwidth]{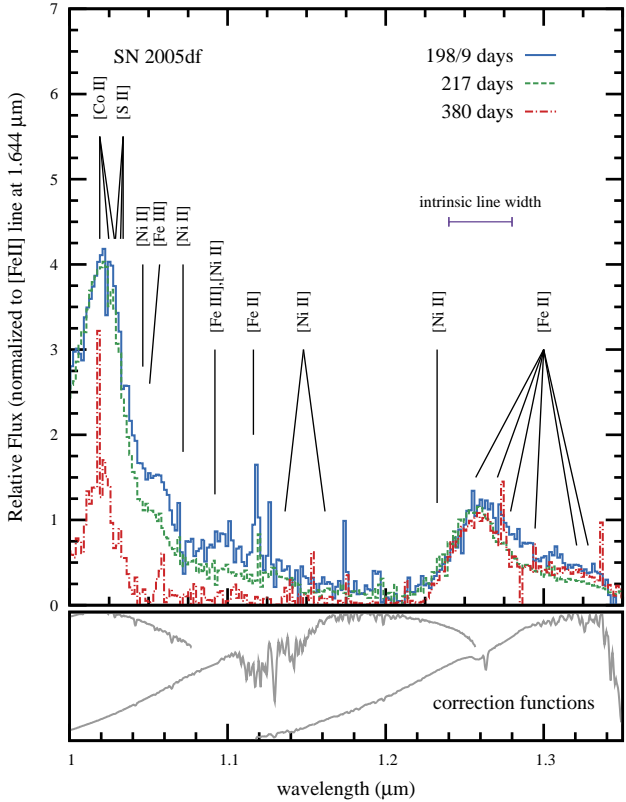}
\caption{Same as Figure~\ref{fig:sci_6-8} but for the region $1.1-1.35\unit{\mu m}$.
Note that all of the features are heavily blended.
In the lower plot, the relative correction factors are given for the atmosphere and the detector response for orders 7, 
6, and 5.
}
\label{fig:sci_5}
\end{figure}

\begin{figure}[h]
\centering
\includegraphics[width=0.5\textwidth]{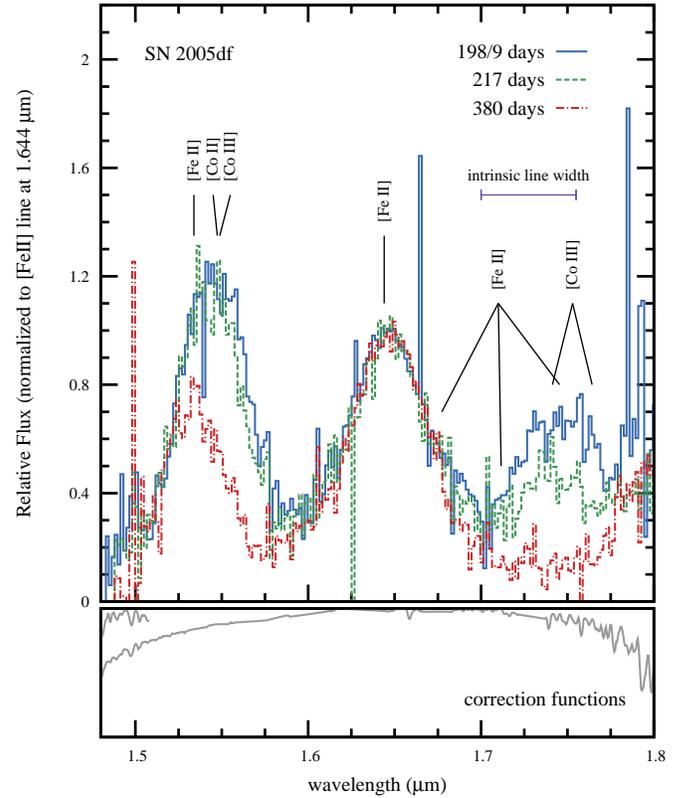}
\caption{Same as Figure~\ref{fig:sci_6-8} but for the region $1.48-1.8\unit{\mu m}$.
In the lower plot, the relative correction factors are given for the atmosphere and the detector response for orders 5 
and 4.
Note the small variability of the correction factor between $1.6-1.7\unit{\mu m}$. 
}
\label{fig:sci_4}
\end{figure}

The reduced spectra span a wavelength range of $\approx0.8-2.4\unit{\mu m}$. 
The spectra are dominated by broad emission features formed by line blends.
Lines in the spectra are identified based on the models described in Section~\ref{sec:models}.
Most of the identifiable emission is due to forbidden line transitions of iron and cobalt.
Large regions of atmospheric absorption and low filter transmission obscure the spectra in the ranges 
$1.35-1.48\unit{\mu m}$ and $1.82-1.95\unit{\mu m}$.
The full spectra are split up into regions corresponding roughly to standard band-passes: 
$0.8-1.1\unit{\mu m}$ (see Figure~\ref{fig:sci_6-8}), J at $1.1-1.35\unit{\mu m}$ (see Figure~\ref{fig:sci_5}), H at
$1.48-1.8\unit{\mu m}$ (see Figure~\ref{fig:sci_4}), and K at $1.9-2.4\unit{\mu m}$.
The XD order 3 data are not shown because of the extremely low signal, especially in the $380\unit{day}$ 
observation, which prohibits any clear line identification in that region.

The XD orders 6, 7, and 8 data (see Figure~\ref{fig:sci_6-8} and Table~\ref{tab:line-id_6-8}) have features that are 
blends of many lines.
Strong features are present at $\approx0.86,0.89,0.95$ and $1.02\unit{\mu m}$, which are dominated by forbidden  
transitions of the iron-group elements and sulfur.
In addition, many forbidden and allowed transitions contribute both to the features and to the overall 
``quasi-continuum'' flux.
The spectral evolution reflects the transition from a cobalt- to an iron-dominated regime.
We note that this region poses some problems related to the atmospheric corrections.
The spectra of the telluric standard star flux reference, an A0V star \citep{kurucz93}, includes features of strong and 
narrow Paschen lines of hydrogen.
In combination with the low wavelength resolution in the telluric reference spectrum discussed in
Section~\ref{sec:obs-red}, the correction produces artifacts in the reduced spectra of SN~2005df.    
In the lower section of Figure~\ref{fig:sci_6-8}, we show the correction factors applied for the three orders that make 
up this wavelength region.
The overall sharp-edged shape of the strongest feature at $0.9-1.0\unit{\mu m}$ is likely caused by the rapid 
variations of the correction factors.
Similarly, the narrow strong line at $\approx 0.951\unit{\mu m}$ is likely artificial.
Its width corresponds to the detector resolution, $\approx 400\unit{km}\unit{s^{-1}}$, and coincides in wavelength 
and size with a feature in the correction factors. 

\begin{deluxetable}{lccc}
\tablecaption{Line identification for SN~2005df ($0.85-1.1\unit{\mu m}$)}
\tablecolumns{4}
\tablewidth{0in}
\tablehead{
\colhead{$\lambda$ ($\unit{\mu m})$}&\colhead{Species}&\colhead{Term}&\colhead{J -- J'}}
\startdata
0.85020&[\ion{Ni}{3}]&\ce{{}^3F} -- \ce{{}^1D}&2 -- 2\\
0.86193&[\ion{Fe}{2}]&\ce{a^4F} -- \ce{a^4P}&9/2 -- 5/2\\
0.87312&[\ion{Fe}{3}]&\ce{a^3P} -- \ce{{}^3D}&2 -- 3\\
0.88406&[\ion{Fe}{3}]&\ce{a^3P} -- \ce{{}^3D}&2 -- 2\\
0.88944&[\ion{Fe}{2}]&\ce{a^4F} -- \ce{a^4P}&7/2 -- 3/2\\
0.90544&[\ion{Fe}{2}]&\ce{a^4F} -- \ce{a^4P}&7/2 -- 5/2\\
0.90711&[\ion{S}{3}]&\ce{{}^3P} -- \ce{{}^1D}&1 -- 2\\
0.92291&[\ion{Fe}{2}]&\ce{a^4F} -- \ce{a^4P}&5/2 -- 3/2\\
0.92701&[\ion{Fe}{2}]&\ce{a^4F} -- \ce{a^4P}&3/2 -- 1/2\\
0.93357&[\ion{Co}{2}]&\ce{a^3F} -- \ce{b^3F}&4 -- 3\\
0.94467&[\ion{Fe}{3}]&\ce{{}^H} -- \ce{a^1G}&5 -- 4\\
0.95332&[\ion{S}{3}]&\ce{{}^3P} -- \ce{{}^1D}&2 -- 2\\
0.96113&[\ion{Fe}{3}]&\ce{{}^H} -- \ce{a^1G}&4 -- 4\\
0.96418&[\ion{Co}{2}]&\ce{a^3F} -- \ce{b^3F}&3 -- 2\\
0.97045&[\ion{Fe}{3}]&\ce{{}^3H} -- \ce{{}^1I}&6 -- 6\\
1.0191&[\ion{Co}{2}]&\ce{a^3F} -- \ce{b^3F}&4 -- 4\\
1.0248&[\ion{Co}{2}]&\ce{a^3F} -- \ce{b^3F}&3 -- 3\\
1.0280&[\ion{Co}{2}]&\ce{a^3F} -- \ce{b^3F}&2 -- 2\\
1.0290&[\ion{S}{2}]&\ce{{}^2D^{o}} -- \ce{{}^2P^{o}}&3/2 -- 3/2\\
1.0323&[\ion{S}{2}]&\ce{{}^2D^{o}} -- \ce{{}^2P^{o}}&5/2 -- 3/2\\
1.0339&[\ion{S}{2}]&\ce{{}^2D^{o}} -- \ce{{}^2P^{o}}&3/2 -- 1/2\\
1.0463&[\ion{Ni}{2}]&\ce{{}^2F} -- \ce{{}^4P}&7/2 -- 5/2\\
1.0507&[\ion{Fe}{3}]&\ce{a^3P} -- \ce{{}^3D}&0 --1 \\
1.0718&[\ion{Ni}{2}]&\ce{{}^2D} -- \ce{{}^4F}&5/2 -- 7/2\enddata
\label{tab:line-id_6-8}
\end{deluxetable}

The XD order 5 data (see Figure~\ref{fig:sci_5} and Table~\ref{tab:line-id_5}) corresponds roughly to J  band.
In addition to the emission feature at $1.02\unit{\mu m}$ described above, a strong feature is apparent at 
$1.28\unit{\mu m}$.
Because the spectra are normalized to the [\ion{Fe}{2}] line at $1.644\unit{\mu m}$ and this $1.28\unit{\mu m}$ 
feature is dominated by blends of [\ion{Fe}{2}], it shows little evolution with time. 
Even though some of these emission lines are quite strong, the proximity of lines and the presence of the 
[\ion{Co}{3}] line, which is decaying over timescales much shorter than that of the iron lines, make this feature 
difficult to use in probing the central chemical structure of the supernova.

\begin{deluxetable}{lccc}
\tablecaption{Line identification for SN~2005df ($1.1-1.35\unit{\mu m}$)}
\tablewidth{0in}
\tablecolumns{4}
\tablehead{
\colhead{$\lambda$ ($\unit{\mu m})$}&\colhead{Species}&\colhead{Term}&\colhead{J -- J'}}
\startdata
1.1163&[\ion{Fe}{2}]&\ce{a^4H} -- \ce{b^2G}&11/2 -- 9/2\\
1.1363&[\ion{Ni}{2}]&\ce{{}^2F} -- \ce{{}^1D}&5/2 -- 3/2\\
1.1616&[\ion{Ni}{2}]&\ce{{}^2D} -- \ce{{}^4F}&3/2 -- 5/2\\
1.2326&[\ion{Ni}{2}]&\ce{{}^2F} -- \ce{{}^4P}&5/2 -- 5/2\\
1.2489&[\ion{Fe}{2}]&\ce{a^6D} -- \ce{a^4D}&7/2 -- 5/2\\
1.2570&[\ion{Fe}{2}]&\ce{a^6D} -- \ce{a^4D}&9/2 -- 7/2\\
1.2707&[\ion{Fe}{2}]&\ce{a^6D} -- \ce{a^4D}&1/2 -- 1/2\\
1.2791&[\ion{Fe}{2}]&\ce{a^6D} -- \ce{a^4D}&3/2 -- 3/2\\
1.2946&[\ion{Fe}{2}]&\ce{a^6D} -- \ce{a^4D}&5/2 -- 5/2\\
1.2981&[\ion{Fe}{2}]&\ce{a^6D} -- \ce{a^4D}&1/2 -- 3/2\\
1.3209&[\ion{Fe}{2}]&\ce{a^6D} -- \ce{a^4D}&7/2 -- 7/2\\
1.3281&[\ion{Fe}{2}]&\ce{a^6D} -- \ce{a^4D}&3/2 -- 5/2\enddata
\label{tab:line-id_5}
\end{deluxetable}

The XD order 4 data and the LS data (see Figure~\ref{fig:sci_4} and Table~\ref{tab:line-id_4}) correspond
roughly to H band.
The dominant emission features are at $1.54,1.64,1.73$, and $1.82\unit{\mu m}$.
The second feature is dominated by one transition, an [\ion{Fe}{2}] line at $1.644\unit{\mu m}$, and shows very little 
evolution in shape over the observed epochs.
Additional [\ion{Fe}{2}] emission lines sit in both the blue and red wings of this main [\ion{Fe}{2}] line, but their 
contributions are far enough out in the wings and low enough in strength that we can actually describe what is 
happening to that single emission line.
The observed line center has a Doppler shift of less than $1200\unit{m}\unit{s^{-1}}$ after correcting for the host
galaxy's radial velocity. 
In contrast, the neighboring features are blends with significant evolution \citep{penney14}. 
The line center of the blended [\ion{Fe}{2}]/[\ion{Co}{3}] feature at $1.54\unit{\mu m}$ shifts toward the red (iron-line 
dominated side) in the $380\unit{day}$ spectrum, which is consistent with the cobalt line fading but the iron line 
remaining strong.
The blended [\ion{Co}{3}] feature at $1.74\unit{\mu m}$ has completely faded in the $380\unit{day}$ spectrum.
The strongly blended [\ion{Fe}{2}]/[\ion{Fe}{3}]/[\ion{Co}{3}]/[\ion{Ni}{2}]/[\ion{Ni}{4}] feature at $1.82\unit{\mu m}$, with wings starting just 
at the red edge of Figure~\ref{fig:sci_4}, does not seem to change between the three epochs, but these lines are 
just at the edge of a low transmission region due to filter overlap and atmospheric corrections. 
The small variation in this feature is indicative of stable isotopes, namely iron and nickel, with the latter being 
observed as [\ion{Ni}{2}] and [\ion{Ni}{4}] in the MIR as isolated lines in \citet{gerardy07} and \citet{telesco14}. 

\begin{deluxetable}{lccc}
\tablecaption{Line identification for SN~2005df ($1.5-1.8\unit{\mu m}$)}
\tablewidth{0in}
\tablecolumns{4}
\tablehead{
\colhead{$\lambda$ ($\unit{\mu m})$}&\colhead{Species}&\colhead{Term}&\colhead{J -- J'}}
\startdata
1.5339&[\ion{Fe}{2}]&\ce{a^4F} -- \ce{a^4D}&9/2 -- 5/2\\
1.5474&[\ion{Co}{2}]&\ce{a^5F} -- \ce{b^3F}&5 -- 4\\
1.5488&[\ion{Co}{3}]&\ce{a^2G} -- \ce{a^2H}&9/2 -- 9/2\\
1.6440&[\ion{Fe}{2}]&\ce{a^4F} -- \ce{a^4D}&9/2 -- 7/2\\
1.6773&[\ion{Fe}{2}]&\ce{a^4F} -- \ce{a^4D}&7/2 -- 5/2\\
1.7116&[\ion{Fe}{2}]&\ce{a^4F} -- \ce{a^4D}&5/2 -- 3/2\\
1.7413&[\ion{Co}{3}]&\ce{a^2G} -- \ce{a^2H}&9/2 -- 11/2\\
1.7454&[\ion{Fe}{2}]&\ce{a^4F} -- \ce{a^4D}&3/2 -- 1/2\\
1.7643&[\ion{Co}{3}]&\ce{a^2G} -- \ce{a^2H}&7/2 -- 9/2\\
1.8099&[\ion{Fe}{2}]&\ce{a^4F} -- \ce{a^4D}&7/2 -- 7/2\\
1.8119&[\ion{Fe}{2}]&\ce{a^4D} -- \ce{a^4P}&7/2 -- 5/2\\
1.8203&[\ion{Fe}{3}]&\ce{{}^1I} -- \ce{{}^3H}&6 -- 6\\
1.8214&[\ion{Co}{3}]&\ce{a^4P} -- \ce{a^2P}&3/2 -- 1/2\\
1.8932&[\ion{Ni}{4}]&\ce{{}^4P} -- \ce{{}^2P}&3/2 -- 3/2\\
1.8958&[\ion{Ni}{2}]&\ce{{}^2D} -- \ce{{}^2P}&3/2 -- 3/2\enddata
\label{tab:line-id_4}
\end{deluxetable}

Over the entire spectral range, the extracted spectra show evolution of cobalt emission lines but minimal evolution 
of iron emission lines, consistent with the decay from cobalt to iron.
The region of most interest for our analysis is around $1.5-1.8\unit{\mu m}$, which includes a mixture of both iron 
and cobalt emission lines.
In Section~\ref{sec:model-comp}, the evolution of the [\ion{Fe}{2}] emission line at $1.644\unit{\mu m}$ is used to 
put limits on the central density of the WD and the strength of the magnetic field in the progenitor by comparing with 
models produced by Hoeflich and Penney \citep{penney11,penney14}.

\section{Models}
\label{sec:models}

Based on detailed spherical models for supernovae discussed below, we analyze the spectral features of
SN~2005df and their evolution.
The simulations are consistent with respect to the explosion, light curves, and spectra.
We emphasize that the time evolution of the NIR spectrum follows without additional free parameters, given the 
initial WD model and parameterized properties for nuclear burning.

\subsection{Methods}
\label{sec:model-methods}

For the calculations of explosions and spectra, we use our code for radiation hydrodynamics
\citep[HYDRA;][]{hoeflich90,hoeflich95,hoeflich02b,hoeflich09} which solves for the hydrodynamics using the 
explicit Piecewise Parabolic Method \citep[PPM;][]{colella84}, detailed nuclear and atomic networks 
\citep{kurucz94-CD20,hoeflich95,seaton05,cyburt10}, transport for low-energy and gamma-transport photons and 
positrons by variable Eddington Tensor solvers and Monte Carlo Methods 
\citep{stone92,hoeflich93,mihalas99,hoeflich02b,hoeflich09,penney14}.
For this study, atomic data for forbidden line transitions have been updated using \citet{quinet96FeII}, \citet{quinet96FeIII}, \citet{quinet98CoII}, \citet{de10}, \citet{friesen14}, \citet[{\it NIST Atomic Spectra Database};][]{NIST}, 
and \citet[{\it The Atomic Line List};][]{ALL}.

\subsection{Model Selection Criteria}
\label{sec:model-select}

In the simulations, we use a spherical, delayed detonation model.
Spherical geometry implies suppression of mixing during the deflagration phase.
Varying the amount of burning prior to the DDT produces a wide range of values for \ce{{}^{56}Ni} mass and the 
corresponding brightness.
It shifts the characteristic chemical pattern in velocity space (see Figure~3 in \citealt{hoeflich02b}).
The explosion models are based on a WD of solar metallicity with a main sequence mass of $7\unit{M_{\sun}}$, 
7p0z22.
Within spherical DDT-models, the amount of \ce{{}^{56}Ni} produced depends mainly on the amount of burning prior 
to the DDT.
This class of models have previously been used with success in reproducing the optical light curves, IR light 
curves, and spectra of a Branch-normal and several sub-luminous SNe~Ia, in addition to the statistical properties of 
the SNe~Ia class 
\citep{hoeflich02,howell06,marion06,quimby07,hoeflich10,maund10a,patat12,marion09,sim13,dessart14}. 
Guided by the 5p0z22-series of models in \citet{hoeflich02} with $\rm{M_V}$ between $-17.21 $ to $-19.35$, 
we choose a model that produces about $0.60\unit{M_{\sun}}$ of \ce{{}^{56}Ni}, which is consistent with 
SN~2005df.
For our WD, 7p0z22, using an initial central density of $0.9 \times 10^9 g/cm^3$, the model peaks at an 
absolute magnitude of $\rm{M_B}=-19.29\unit{mag}$ and has decline ratios of  $\Delta m_{15}(B)=1.14\unit{mag}$ 
and $\Delta m_{15}(V)=0.61\unit{mag}$.
These values are comparable to the observed values for SN~2005df of $\Delta m_{15}(B)=1.2$, 
$\Delta m_{15}(V)=0.63\unit{mag}$, and $\rm{M_B}=-19.23(\pm0.10)$, assuming a distance modulus of 
$\mu=31.81$ as in \citet{milne10}.

For this study, the initial central density of the WD, $\rho_c$ in units of $10^9 g/cm^3$, and initial magnetic field 
have been varied between $0.5-4.0$ and $10^0-10^{9}G$, respectively, to provide a range of models to analyze 
the NIR spectra of SN~2005df. 
Note that the resulting light curve variations are small, $\delta(\rm{M_V})\approx\pm0.04\unit{mag}$, 
$\delta(\Delta m_{15}(B))\approx\pm0.07$, and $\delta(\Delta m_{15}(V))\approx\pm0.03$ 
\citep{hoeflich06a,hoeflich10}, and can be ignored.

\section{Results}
\label{sec:results}

\subsection{Effects of Central Density}
\label{sec:cen-den}

\begin{figure}[h]
\centering
\includegraphics[width=0.5\textwidth]{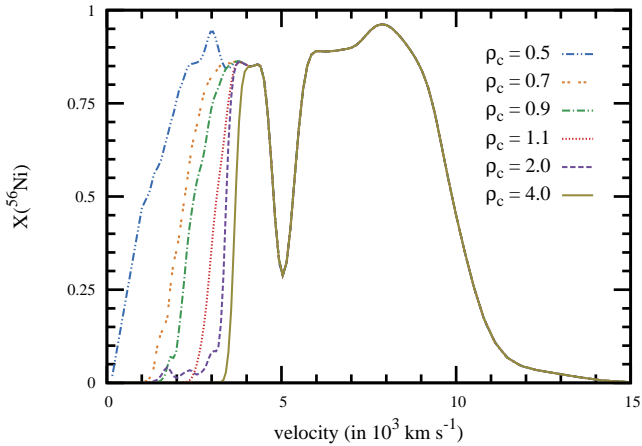}
\caption{The abundance of \ce{{}^{56}Ni} at $t=0$ as a function of the expansion velocity for models with various 
initial central densities $\rho_c$ of the WD (in $10^9\unit{g}\unit{cm^{-3}}$).
The lack of \ce{{}^{56}Ni} is due to electron capture which shifts the NSE to stable isotopes of iron group elements.
Thus, the size of the \ce{{}^{56}Ni} hole increases with $\rho_c$.
The dip at $5000\unit{km}\unit{s^{-1}}$ is an artifact of spherical delayed detonation models and caused by the
strong reflection wave produced by the DDT.
}
\label{fig:ni_dist}
\end{figure}

Because all of the reference models are constructed with a very similar density at which the DDT occurs, the 
abundance profiles of \ce{{}^{56}Ni} look identical in the outer regions of velocity-space, as can be seen in 
Figure~\ref{fig:ni_dist}.
However, the interior profile varies dramatically, with radioactive nickel providing less and less of a low-velocity
contribution with increasing central density.
We note that this velocity profile does not mean the emission line widths will be on the order of 
$10,000\unit{km}\unit{s^{-1}}$.
Assuming a spherical distribution of material, the line widths will actually be significantly narrower because of
projection effects.

\begin{figure}[h]
\centering
\includegraphics[width=0.5\textwidth]{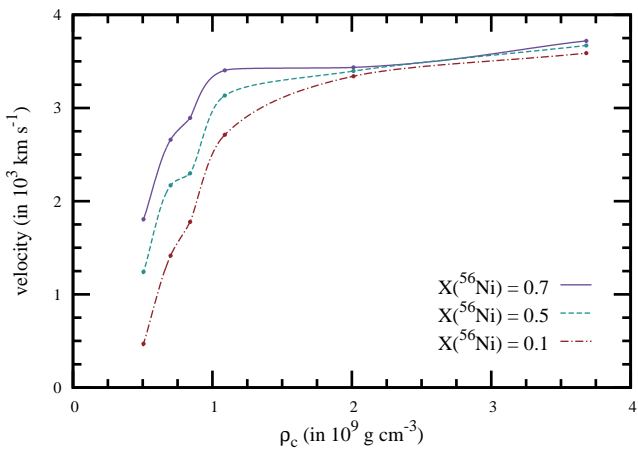}
\caption{Size of the \ce{{}^{56}Ni} hole in velocity space as a function of $\rho_c$.
Here the edge of the hole is defined by abundance levels of $0.7$, $0.5$, and $0.1$.
}
\label{fig:v-rho}
\end{figure}

As the central density increases, there is a larger fraction of material at high enough density during the deflagration 
for electron capture to occur and produce sizable amounts of stable \ce{{}^{58}Ni} 
\citep{hoeflich06a}.
This ``hole'' in the radioactive material does not contribute to the emission features in the nebular spectrum until 
very late times when the positrons have become non-local.
Figure~\ref{fig:v-rho} shows the size of the \ce{{}^{58}Ni} hole increasing with $\rho_c$, and this corresponds to a broadening 
of the nickel, cobalt, and iron emission lines.

\subsection{Benefits of the H-band Region}
\label{sec:H-band}

As can be seen from the SN~2005df spectra in Figures~\ref{fig:sci_6-8}-\ref{fig:sci_4}, there are significant 
overlapping emission lines contributing to most of the features seen in the NIR.
The ``cleanest'' feature in the entire observed region is the [\ion{Fe}{2}] line at $1.644\unit{\mu m}$, whose closest 
satellite line with moderate strength is $0.333\unit{\mu m}$ away from line-center.
The entire H-band region from $1.5-1.8\unit{\mu m}$ can effectively be modeled using just the contributions from 
the strongest emission lines listed in Table \ref{tab:line-id_4}.
In this analysis, we will find which of the references models best fits with the data by considering only the 
$1.644\unit{\mu m}$ line.
The other NIR features come out naturally.

\subsection{Model Comparison}
\label{sec:model-comp}

By $200\unit{days}$ after the explosion the energy deposition transitions from gamma-dominated to 
positron-dominated.
Unlike gamma rays, whose path is not affected by a magnetic field embedded in the ejecta, the 
positrons are very much dependent on whether or not there is a magnetic field present and the strength of that field.
With no magnetic field, positrons deposit energy locally early on, and we would expect any central 
region of \ce{{}^{58}Ni} to not contribute to the observed emission lines, providing a ``flat-topped'' or ``stubby'' 
line profile.
However, with time, positrons are able to travel throughout the central non-radioactive region, and we transition 
from seeing a flat-topped profile to a peaked profile, with the line evolving most quickly in the wings where the 
higher-velocity material is able to become non-local first.
A more in-depth discussion of line evolution with magnetic fields is presented in Section~\ref{sec:B-effect}.

\begin{figure}[h]
\centering
\includegraphics[width=0.5\textwidth]{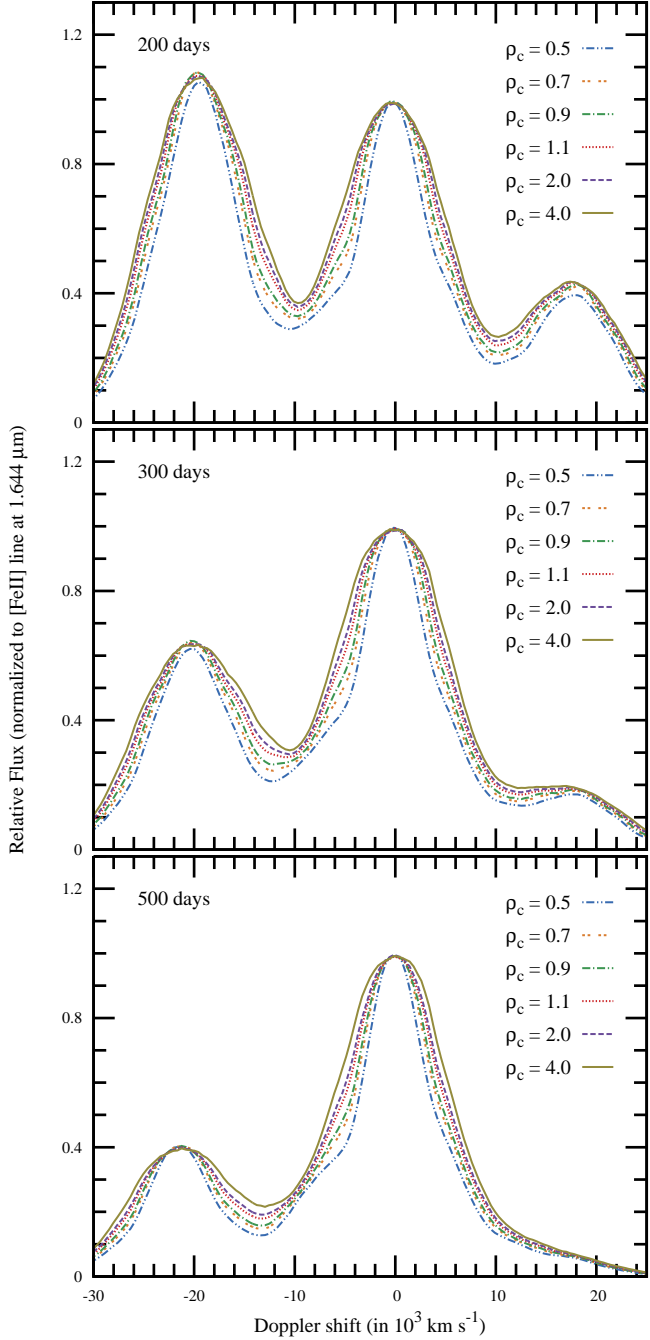}
\caption{Time evolution of the [\ion{Fe}{2}] emission line at $1.644\unit{\mu m}$ in velocity space of our reference 
models with $B=10^0\unit{G}$.
In wavelength this region spans $1.48-1.78\unit{\mu m}$.
The strong neighboring features on the blue and red side are produced by blends of [\ion{Fe}{2}]/[\ion{Co}{3}] and 
[\ion{Co}{3}], respectively (see Table \ref{tab:line-id_4}).
}
\label{fig:model_comp_200}
\end{figure}

\subsection{Central Density Fitting}
\label{sec:cen-den-fit}

\begin{figure}[h]
\centering
\includegraphics[width=0.5\textwidth]{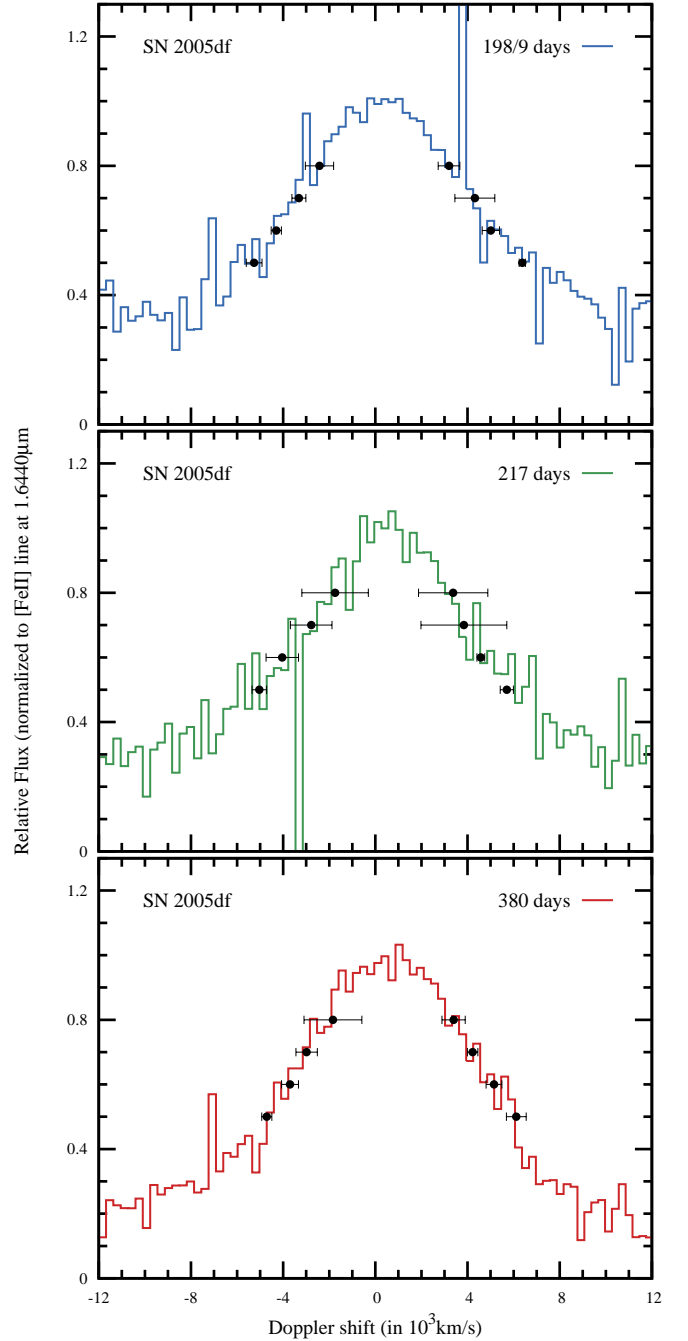}
\caption{Observed spectra at $198/199\unit{days}$, $217\unit{days}$, and $380\unit{days}$ for the 
[\ion{Fe}{2}] $1.644\unit{\mu m}$ line. 
The plot shows the velocity approximations at different heights ($0.5\leq H\leq 0.8$) overlaid on the actual data.
These values are the ones used to determine line width for the observational data.
We limit the range for $H$ because the wings of this feature are contaminated by neighboring iron and cobalt 
emission lines.
}
\label{fig:data}
\end{figure}

As discussed in Section~\ref{sec:cen-den}, the distribution of iron group elements in the central regions depends 
on the initial central density of the WD. 
Therefore, the shape of the line profile, specifically the line width, is sensitive to measuring $\rho_c$.
Our procedure for finding the line width ($LW$) of the $1.644\unit{\mu m}$ line is as follows:
\begin{enumerate}
\item subtract the free-free continuum produced by the models, re-normalizing to the $1.644\unit{\mu m}$ peak, to 
give the line profile a zero-point;
\item pick fixed $y$-value points (denoted $H$ in plots) in increments of $0.1$ or smaller, as signal-to-noise of the 
data allow;
\item for each of the chosen $y$-values, the corresponding velocity $x$-values are obtained using a linear 
regression through adjacent data points, with multiple resolution elements used in the case of the data to decrease 
the effects of noise;
\item for each epoch and magnetic field strength, measure $LW$ from the $+x$-value to the $-x$-value for each of 
the chosen $y$-values.
\end{enumerate}
Because of this treatment, line evolution is determined only by drop-out in the wings or core and does not include 
changes in the total line flux.
This representation allows for direct comparison to our observational data, in which total flux could not be  
measured.
Figure~\ref{fig:data} shows the observational data and the velocity $x$-values we obtain for each $y$-value.

The continua produced by the reference models are at levels of $9\%$, $7\%$, and $2\%$ of the line-height for the 
$198/9\unit{day}$, $217\unit{day}$ and $380\unit{day}$ epochs, respectively.
These modeled continua are also taken to zero-point the SN~2005df data.
However, this zero-point uncertainty may add additional error to our measurements that will systematically increase 
the measured $LW$.
A maximum possible continuum can be determined using the minimum observed flux in the corresponding  
wavelength range (Figure~\ref{fig:sci_4}).
Using this method, the systematic errors in the continua are less than $15\%$, $16\%$, and $12\%$ for the 
$198/9\unit{day}$, $217\unit{day}$, and $380\unit{day}$ observations, respectively, before the modeled continua 
are removed.
The systematic error estimates in $LW$ are obtained as follows: use the continuum error to determine the 
maximum adjusted $H'$; use the original $H$ values to interpolate new velocity values at $H'$; obtain a new 
estimate for $LW'$, where the systematic uncertainties are given by $LW'-LW$.
For our observations, comparing $H=0.6$ and $H'$, we get $H'=0.624$, $0.636$, and $0.64$ for the three epochs.
The systematic error due to continuum uncertainty will affect the line width at a maximum level of 
$600\unit{km}\unit{s^{-1}}-700\unit{km}\unit{s^{-1}}$, with the $217\unit{day}$ observation having the largest 
possible effect.

\begin{figure}[h]
\centering
\includegraphics[width=0.5\textwidth]{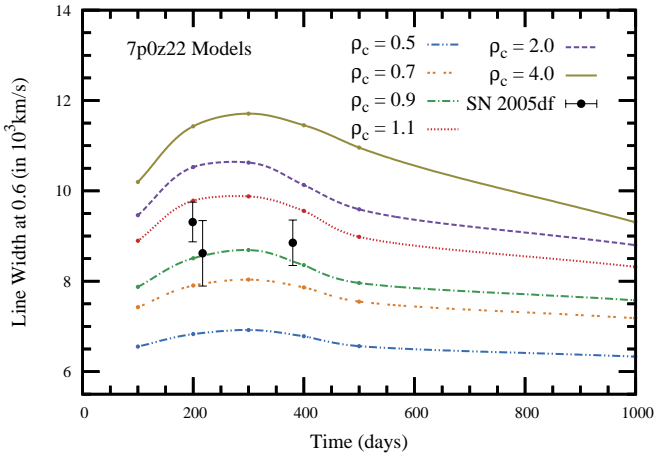}
\caption{Comparison between $LW$ observed for SN~2005df and our reference models at
$H=0.6$ for the [\ion{Fe}{2}] emission line at $1.644\unit{\mu m}$. 
The errors bars given are produce by the noise in the data (see text). 
Good agreement with the SN~2005df data can be obtained for models with $\rho_c$ between $0.9$ and $1.1$.
}
\label{fig:line-width}
\end{figure}

One of the advantages of the method we have developed is that there is not one specific measurement that must be 
made.
The line width can be measured at a variety of different $H$-values to ensure that the results found are consistent.
Noise in the data in small wavelength regions will not affect the overall result.
The $y$-value of $0.6$ has been chosen for display in Figures~\ref{fig:line-width} and \ref{fig:line-width_zoom} 
even though each $y$-value we tested produces very similar results. 
At $0.6$ the level of contamination from the neighboring satellite lines to $1.644\unit{\mu m}$ [\ion{Fe}{2}] line, 
which increases with decreasing $H$, is small.
Similarly, the uncertainty when finding the corresponding velocity $x$-values, which increase with increasing $H$, 
is small because of the slope of the line in that region.

\begin{figure}[h]
\centering
\includegraphics[width=0.5\textwidth]{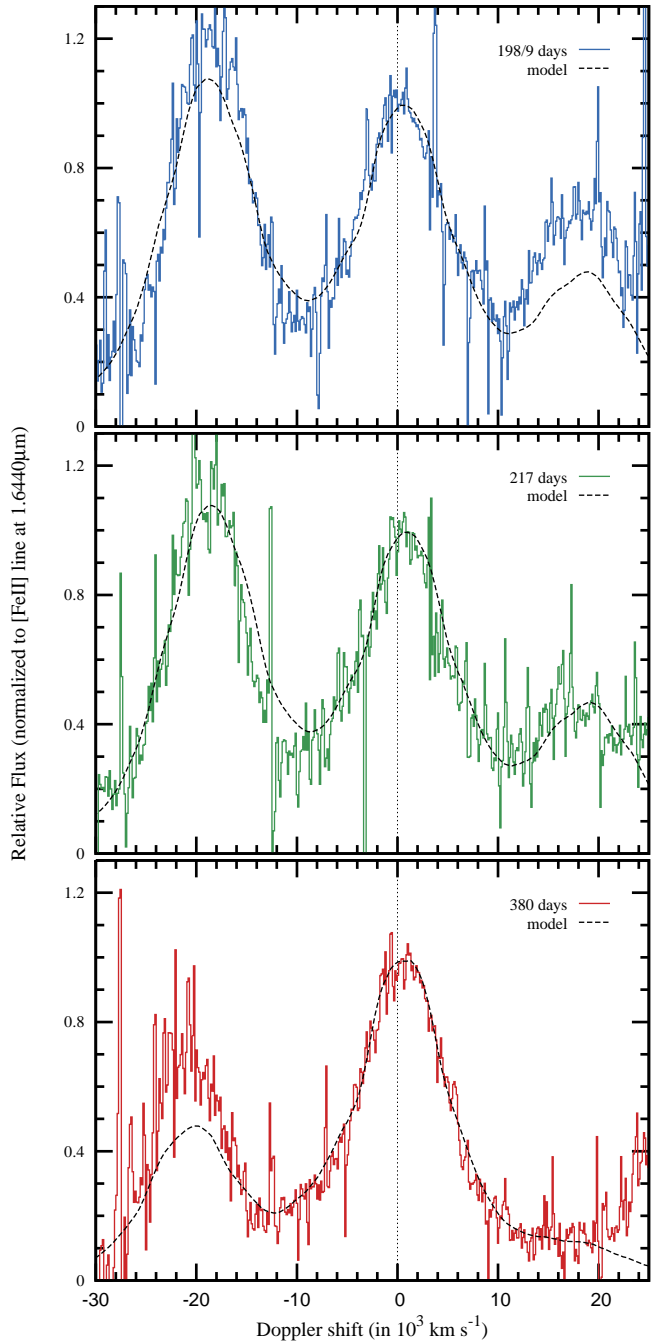}
\caption{Comparison between spectra of  SN~2005df  and  the  reference model with  
$\rho_c=0.9$ and $B=10^0\unit{G}$.
Note the discrepancies of the relative flux in the neighboring features which may be attributed to uncertainties in the 
continuum or the model atoms (see text).
}
\label{fig:comp_H}
\end{figure}

Using $LW$ to determine the initial central density of the WD for SN~2005df, we find that 
$\rho_{c}({\rm SN~2005df})=0.9(\pm0.2)$ (in $10^9\unit{g}\unit{cm^{-3}}$).
The systematic error due to continuum uncertainties is just over the $1\sigma$-level but can only make $LW$ 
narrower and, therefore, the central density lower.
This central density is high enough that SN~2005df would have a moderately-sized region of stable \ce{{}^{58}Ni} 
at low velocity, however not enough to produce the characteristic ``flat-topped'' line profiles seen in SN~2003du 
and SN~2003hv.
The best-fit reference model is shown plotted with the SN~2005df data in Figure~\ref{fig:comp_H}.
The line-center for the [\ion{Fe}{2}] line at $1.644\unit{\mu m}$ is shifted by less than $1200\unit{km}\unit{s^{-1}}$.
The main feature is very well reproduced in line profile both in the core and wings of the line. 
There are some discrepancies in the relative fluxes for the neighboring features around 
$-20,000\unit{km}\unit{s^{-1}}$ (corresponding to $1.54\unit{\mu m}$) and $18,000\unit{km}\unit{s^{-1}}$ 
(corresponding to $1.74\unit{\mu m}$).
However, these may be attributed to the treatment of super-levels in the reference models or in the simplistic 
treatment of continuum subtraction.

\subsection{Effects of Magnetic Fields}
\label{sec:B-effect}

By looking at the initial line profile and the evolution, we can put a lower limit on the strength of a magnetic field 
embedded in the SN ejecta.
Large magnetic fields embedded in the SN ejecta keep the energy deposition of the positrons local longer than in a 
$B=0\unit{G}$ model, because the positron path-length increases dramatically.
In order to probe this effect, we have included a range of field strengths for a magnetic field embedded in the SN 
ejecta:  $B=10^0\unit{G}$, $10^3\unit{G}$, $10^4\unit{G}$, $10^6\unit{G}$, and $10^9\unit{G}$.

\begin{figure}[h]
\centering
\includegraphics[width=0.5\textwidth]{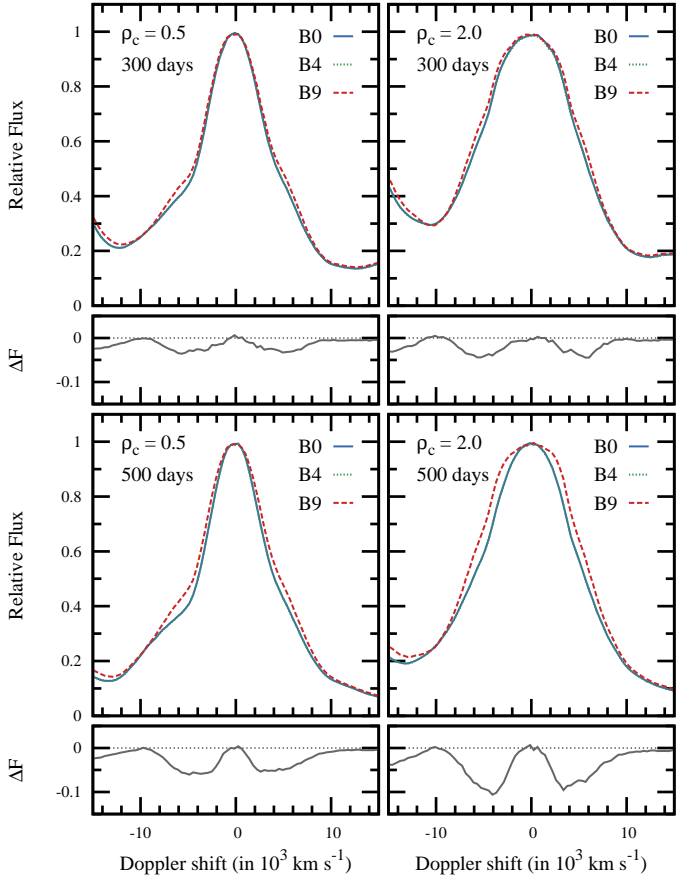}
\caption{Influence of the strength of the magnetic field on the angle-averaged line profile of the $1.644\unit{\mu m}$ 
[\ion{Fe}{2}] line at $300\unit{days}$ and $500\unit{days}$.
Below each plot, the difference in line profile is shown between $B=10^0\unit{G}$ (B0) and $B=10^9\unit{G}$ (B9).
Positrons are responsible for the energy input during late times $t$. 
At $300\unit{days}$, the influence of $B$ is smaller than a shift of $1000\unit{km}\unit{s^{-1}}$ in all of the reference 
models.
At $500\unit{days}$, the influence of $B$ on the core of the line ($H=0.6$) for the low and high density model is 
$\approx750\unit{km}\unit{s^{-1}}$ and $\approx1750\unit{km}\unit{s^{-1}}$, respectively. 
This puts the magnetic field effect close to the resolution limit of our measurements for SN~2005df, assuming a 
relatively low initial central density as found in Section \ref{sec:cen-den-fit}.
Note that the wing is determined by  high velocity \ce{{}^{56}Ni} justifying our choice of $H=0.6$ to measure 
$\rho_c$. 
}
\label{fig:B-size}
\end{figure}

\begin{figure}[h]
\centering
\includegraphics[width=0.5\textwidth]{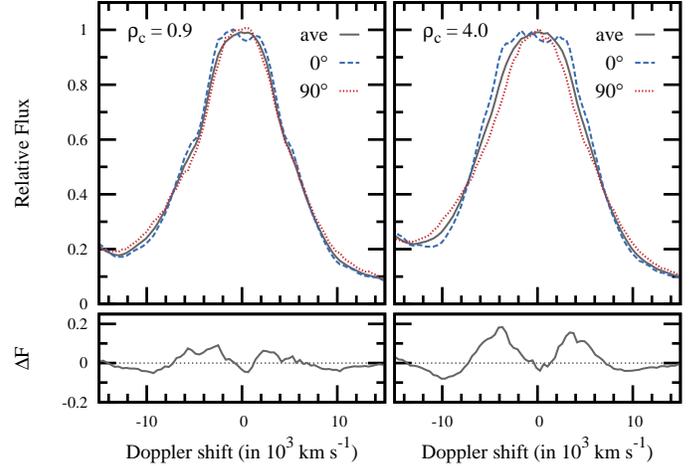}
\caption{Same as Figure \ref{fig:B-size} but showing the influence of the observation angle.
Orientation effects remain small for $\rho_c=0.9$ even at $500\unit{days}$.   
}
\label{fig:B-orientation}
\end{figure}

For this analysis, the model data are averaged over angle for each epoch and progenitor field strength, although 
discussion of the effect of orientation is presented below and shown in Figure~\ref{fig:B-orientation}.
In order to look at the effects of magnetic field strength, a low-$\rho_c$ and a high-$\rho_c$ case are shown in 
Figure~\ref{fig:B-size} at $300\unit{days}$ (top panel) and $500\unit{days}$.
Although the energy deposition is dominated by positrons at $300\unit{days}$, the positrons are still localized and 
even the most extreme magnetic field, $B=10^9\unit{G}$, only affects the flux to a minor extent.
By $500\unit{days}$, the magnetic field effect on the flux is within a measurable range for the high-$\rho_c$ 
spectrum, with the extreme $B=10^9\unit{G}$ spectrum retaining its ``flat-topped'' profile.

The morphology of the magnetic field can also affect the line profile shape and evolution; however, only models 
with a dipole field are considered here \citep{penney14}.
Additionally, there is an effect of the viewing angle with respect to the orientation of the magnetic field's dipole, 
which impacts the line profile shape and evolution.
In most of the reference models, the difference between the flux when viewed from a polar orientation versus an 
equatorial orientation is smaller than the resolution limit for our SN~2005df observations and will be negligible for 
our purposes.
In cases of large central densities, however, the difference between flux for the normalized line can be as large as 
$\Delta F=0.2$, and this effect should be considered along with other variations in the line profile.
Two cases are shown in Figure~\ref{fig:B-orientation}, both at $500\unit{days}$: 
$\rho_c=0.9\times10^9\unit{g}\unit{cm^{-3}}$ with $B=10^9\unit{G}$ and 
$\rho_c=4.0\times10^9\unit{g}\unit{cm^{-3}}$ with $B=10^9\unit{G}$.

\begin{figure}[h]
\centering
\includegraphics[width=0.5\textwidth]{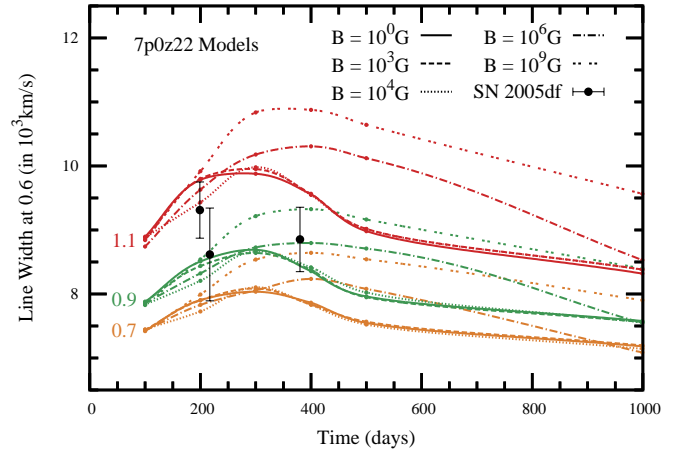}
\caption{Same as Figure \ref{fig:line-width} but for a reduced sample with magnetic field effects included.
For low $\rho_c$, $B$ effects remain small.
The late time observation of SN~2005df favors $B\approx10^6\unit{G}$ by a $1\sigma$ level only.
} 
\label{fig:line-width_zoom}
\end{figure}

A rough cut of central densities can be made using line widths.
However, even by $200-300\unit{days}$, a strong magnetic field keeps $LW$ larger when compared to the 
$B=0\unit{G}$ model.
With the SN~2005df observations, the two effects are indistinguishable because of the noise in the data and the 
limited epochs of observation, as can be seen in Figure~\ref{fig:line-width_zoom}.
A non-zero magnetic field is only favored by a $1\sigma$-level.
Figure~\ref{fig:line-width_zoom} shows that an observation at $400\unit{days}$ post-explosion or later, in addition 
to an early nebular phase observation, is needed in order to obtain both $\rho_c$ and magnetic field strength.

\subsection{Overall Spectral Comparison for the Wavelength Range $0.8-1.35\unit{\mu m}$}
\label{sec:nir-comp}

\begin{figure}[h]
\centering
\includegraphics[width=0.5\textwidth]{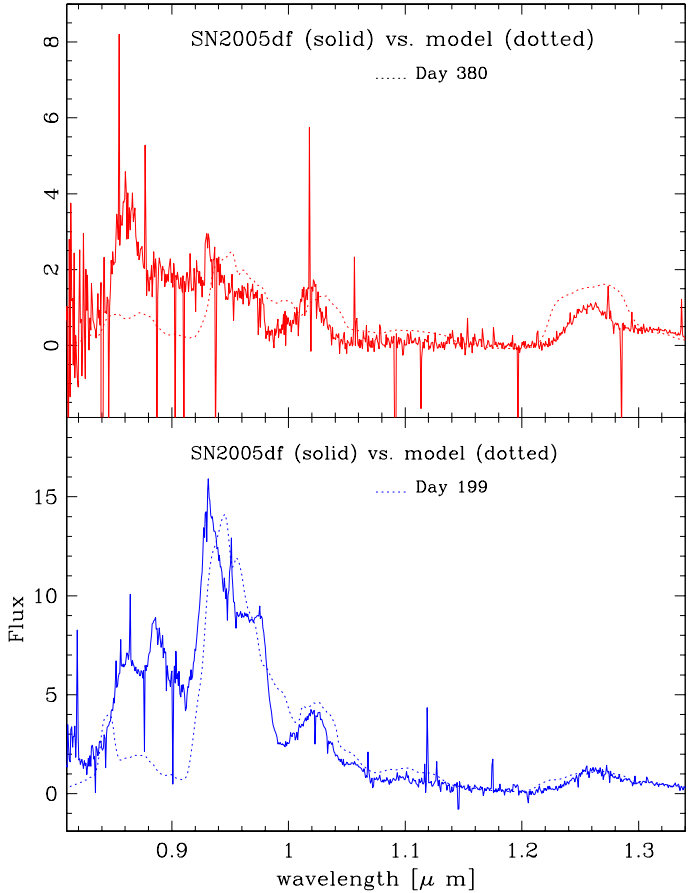}
\caption{Same as Figure \ref{fig:comp_H} but for the short wavelength range.
The fluxes have been normalized to the $1.644\unit{\mu m}$ [\ion{Fe}{2}] line.
}
\label{fig:comp_X}
\end{figure}

In this section we compare the short wavelength region, $0.8-1.35\unit{\mu m}$, of SN~2005df with our 
best-fit reference model (Figure \ref{fig:comp_X}).
The fluxes have been normalized to the $1.644\unit{\mu m}$ [\ion{Fe}{2}] line without further tuning.
Most of the features seen in SN~2005df can be reproduced, which allows for the identification of the main spectral 
features as given in Tables \ref{tab:line-id_6-8} and \ref{tab:line-id_5}. 
At both $200\unit{days}$ and $400\unit{days}$, the spectra are dominated by iron group elements, and evolution of 
the spectra can be understood by the transition from a cobalt-dominated to an iron-dominated regime.
This is evident by the rapid evolution of the broad feature at $0.95\unit{\mu m}$ compared to the slow 
changes in the iron-dominated features at $0.86\unit{\mu m}$, $1.03\unit{\mu m}$, and $1.26\unit{\mu m}$, which 
are produced by blends of [\ion{Fe}{2}], [\ion{S}{2}], and [\ion{Fe}{2}], respectively.
Both the width and flux ratios of features are reproduced in the model.
We note, however, that each of the main features is strongly blended, emphasizing the uniqueness of the 
[\ion{Fe}{2}] emission line at  $1.644\unit{\mu m}$ used in the previous sections for central density and magnetic 
field strength analyses.

By $200\unit{days}$ after explosion, the envelope is mostly transparent.
Most of the radioactive \ce{{}^{56}Ni} and \ce{{}^{56}Co} has decayed to \ce{{}^{56}Fe}, and stable nickel and 
iron are fully exposed.
The models show, and we want to emphasize, that normalization of spectra to this line allows us to separate whether 
features are dominated by stable or unstable isotopes.

Some discrepancies are clear.
The flux between below $0.92\unit{\mu m}$ is too low by a factor of 3, and a narrow feature is present at 
$0.95\unit{\mu m}$ with a width corresponding to the instrumental resolution of $\approx400\unit{km}\unit{s^{-1}}$.  
The problem may be related to the use of super-levels and the lack of atomic data with the reference model.
However, tests with various cross sections from literature produce changes of only $20-30\%$.
Alternatively, the inconsistency may be related to the atmospheric corrections during data reduction.
As discussed in Section~\ref{sec:obs-red}, we use the fluxes calibrated by the energy distribution of an A0V 
standard star with a resolution of $20\unit{\AA}$ to $50\unit{\AA}$, which is larger than the detector resolution in
the wavelength range shown in Figure~\ref{fig:comp_X}.
Additionally, A0V stars have strong lines in the Paschen series in this region, and the correction factors vary by 
$\approx 2$, which may produce an overcorrection of the SN~2005df flux.

Moreover, a strong, narrow feature at $0.95\unit{\mu m}$ is unexpected based on the models and would imply a 
component with high emissivity in a very localized region, $\approx400\unit{km}\unit{s^{-1}}$, in the envelope. 
In the line list, this feature could be [\ion{S}{3}], but it is difficult to imagine a scenario where sulfur can be excited at 
high levels without a corresponding narrow feature in cobalt.
Therefore, we regard the atmospheric correction as the likely source, especially considering the coincidence of the 
feature with a Paschen line in the A0V telluric standard star.  
                                                                                                                                                                                              
\subsection{Implications of the Central Density for the Possible Progenitor and Explosion Scenario of SN~2005df} 
\label{sec:rho-m}

\begin{figure}[h]
\centering
\includegraphics[width=0.5\textwidth]{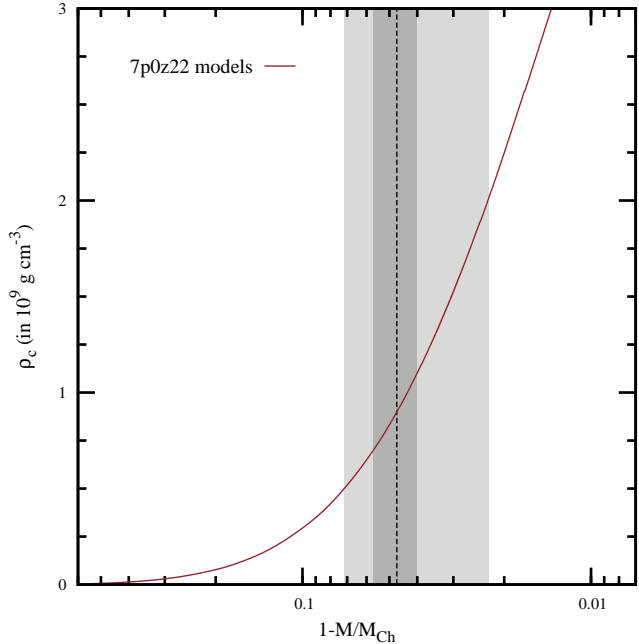}
\caption{Central density $\rho_c$ as a function of the progenitor.
There is a rapid increase of $\rho_c$ towards $\rm M_{Ch}$.
Therefore we show $\rho_C$ as a function of $1-{\rm M_{WD}}/{\rm M_{Ch}}$ (mass fraction away from $\rm M_{Ch}$).
In addition, the best-fit central density for SN~2005df is given, with the dotted line showing the corresponding 
$1-{\rm M_{WD}}/{\rm M_{Ch}}$.
The dark and light gray areas denote the regions with $1\sigma$ and $2\sigma$ probability which span 
$\rho_c=0.7-1.1$ and $\rho_c=0.5-2.0$ (in units of $10^9\unit{g}\unit{cm^{-3}}$), respectively.
}
\label{fig:m-rho}
\end{figure}

As discussed in Section~\ref{sec:intro}, different progenitor systems and explosion scenarios can be distinguished 
by $\rho_c$. 
The pressure in the WD is dominated by a degenerate electron gas.  
With increasing density, this Fermi gas becomes increasingly relativistic degenerate, and $\rho_c$ increases 
rapidly when approaching $\rm M_{Ch}$.
As a consequence, $\rho_c$ is a steep function of the WD mass, $\rm M_{WD}$ (see Figure~\ref{fig:m-rho}).
We note that the equation of state determines the $\rm M_{Ch}$ for the WD and depends on the metallicity, 
composition, and accretion. 
However, in practice, a WD cannot reach this limit because densities get high enough that the time scales for
electron capture become shorter than the hydrodynamical time scales, which reduces pressure in the WD core and 
causes an accretion induced collapse (AIC). 
In Figure~\ref{fig:m-rho}, our reference mass for $\rm M_{Ch}$ is taken when 
$\rho_c\approx 7.0$ (in units of $1\times10^9\unit{g}\unit{cm^{-3}}$) and assumes low accretion rates.

For SN~2005df, we obtain $\rho_c=0.9(\pm0.2)$ (in $10^9\unit{g}\unit{cm^{-3}}$) in Section~\ref{sec:cen-den-fit}.
Our systematic error due to continuum uncertainty is at about the $1\sigma$-level.
Despite the rather large error range in $\rho_c$, the steep dependence of $\rho_c$ on the mass of the WD puts a 
strong constraint on $\rm M_{WD}(SN~2005df)=1.313(\pm0.034)\unit{M_{\sun}}$ with a $2\sigma$ confidence and 
systematic error less than $0.04\unit{M_{\sun}}$.  
Mixing of \ce{{}^{56}Ni} will increase the central emission component, thus decreasing the line width of the 
$1.644\unit{\mu m}$ line and mimicking a lower central density. 
Therefore, $\rho_c({\rm SN~2005df})$ and the corresponding $\rm M_{WD}(SN~2005df)$ should be taken as lower 
limits.
Because dynamical mergers occur in a wide range of masses, they are unlikely progenitors for SN~2005df.

For $\rm M_{Ch}$ scenarios, the explosion is triggered by compressional heating due to increasing mass 
by accretion from the companion star.
To first order,  $\rho_c$ is determined by the balance of this heating with cooling by conduction.
The $\rho_c({\rm SN~2005df})$ found in our analysis implies an upper end for the rate of 
\ce{H}-accretion.
As discussed in Section~\ref{sec:intro}, higher accretion rate can be achieved for accretion of \ce{He} and \ce{C}.
 
Electron capture takes place in the center of WDs when they approach $\rm M_{Ch}$ prior to the explosion, 
during the convective ``smoldering'' phase.
The Urca process, which involves neutrino emission as a consequence of electron capture, was thought to cool the 
WD center and delay the onset of explosive \ce{C} burning \citep{paczynski72,barkat90}.
This would mean that $\rm M_{Ch}$ scenarios should transition into an explosion only at central densities in 
excess of $3\times10^9\unit{g}\unit{cm^{-3}}$.
However, this notion was recently debated.
It was noted that neutrino emission also has the opposite effect of slowing down the convection and thus heating 
the core \citep{stein99}.
Cooling and heating due to the Urca process may cancel out \citep{stein06}, and our results lend support to this 
conclusion.

\section{Conclusions}
\label{sec:conclusions}

We have shown the time evolution of the NIR spectrum of SN~2005df, specifically focusing on the [\ion{Fe}{2}] 
emission line at $1.644\unit{\mu m}$.
We found that this line provides a unique tool to analyze near-infrared spectra.
Normalization to this line allows separation of features produced by stable versus unstable isotopes of iron group 
elements because the envelope is mostly optically thin by day 200.

We have developed a methodology for using late-time NIR spectra of SNe~Ia to determine the initial central density  
and strength of embedded magnetic field of the WD.
Time-series observations in the nebular phase are of great importance because they shed light on the interior 
structure of the burning products in the SN.
There is currently a scarcity of observations showing late-time NIR results for SNe~Ia and time evolution 
observations of the nebular spectra of SNe~Ia.

The [\ion{Fe}{2}] emission line at $1.644\unit{\mu m}$ is relatively ``clean'' compared to most of the NIR region, as 
was shown in Sections~\ref{sec:pre-analysis} and \ref{sec:H-band}, especially when compared with the optical 
spectra of nebular SNe~Ia.
The neighboring features to the $1.644\unit{\mu m}$ line are blended features, with multiple iron and cobalt 
emission lines contributing, and are therefore not as well-suited to analyze the evolution of line profiles, which has 
been presented in this work.
The [\ion{Fe}{2}] $1.644\unit{\mu m}$ emission line is ideal because of the relatively little line 
contamination from nearby features.

An early nebular epoch, somewhere around $100-200\unit{days}$ past explosion, is needed to pin down the 
central density of the WD before the magnetic fields in the ejecta start affecting the line width curves.
As demonstrated by the models, the line widths of the [\ion{Fe}{2}] $1.644\unit{\mu m}$ line converge when 
gamma emission still dominates in the nebular phase and is independent of the magnetic field strength embedded 
in the ejecta.
Multiple epochs in the nebular phase are needed in order to probe the evolution of the [\ion{Fe}{2}] 
$1.644\unit{\mu m}$ emission line and set a lower limit on the magnetic field strength, which plays a role once 
positron emission dominates.

Finally, we have discussed some implications of a low mass WD and put our results into context of progenitor 
systems and explosion scenarios. 
The $\rm M_{WD}(SN~2005df)$ we have obtained is close to $\rm M_{Ch}$, making a violent or dynamical merger 
scenario unlikely because those can trigger explosions in a wide range of masses.
Within the scenario of $\rm M_{Ch}$ explosions, the thermonuclear runaway is triggered by compressional heating 
due to accretion of \ce{H}, \ce{He} or \ce{C} as discussed in the introduction. 
A low central density, such as we found for SN~2005df, requires high accretion rates in $\rm M_{Ch}$ scenarios.
For our models, \ce{H}-accretion would barely be consistent with ignition at our measured $\rho_c$ for SN~2005df, 
even without Urca process cooling \citep{barkat90,stein06}.
Our results for SN~2005df may favor \ce{He}-accretion from a giant companion star in a SD system 
or \ce{C}-accretion from a disrupted WD companion in a DD system. 
Alternatively, SN~2005df may show some central mixing and originate from a WD even closer to $\rm M_{Ch}$.
Note that  MIR spectra, with their unblended \ce{Ni} lines during the first few months when the envelope changes 
from the optically thick to transparent continua, allow for investigation of possible mixing and $\rm M_{Ch}$ mass 
explosions \citep{telesco14}.

The methods we have developed for line width determination of central density and magnetic fields is applicable to 
a large range of SNe~Ia.
Specifically, SN~2014J has very similar light curve characteristics to SN~2005df, with 
${\rm M_B}=-19.19(\pm0.10)$ and $\Delta m_{15}(B)=1.12$ \citep{marion14}, mid-IR spectra 
\citep{gerardy07,telesco14}, and \ce{{}^{56}Ni} mass based on gamma-ray observations 
\citep{churazov14,diehl14,isern14}.
These two SNe appear to be identical in the exterior regions of the explosion but what about the interior regions?
NIR spectra should be of high enough signal-to-noise to determine initial WD central density and magnetic fields.
Differential comparison of SN~2005df and SN~2014J should therefore put new constraints on the progenitor 
system and the explosion scenario.

With the observations of SN~2005df presented in this work, we are able to probe central density of the WD but are 
just on the edge of being able to probe the influence magnetic fields have on the spectra.
Moreover, we need a statistical sample to examine the diversity of SN~Ia and the variety of explosion scenarios that 
all may be realized in nature.
With forthcoming instruments like the James Webb Space Telescope, the Very Large Telescope, the Wide-Field 
Infrared Survey Telescope, the Giant Magellan Telescope, and the Extremely Large Telescope, many more 
observations of SNe~Ia in the nebular phase will be available at better signal-to-noise.
Additionally, the available atomic data needed for modeling is increasing and improving.
This type of analysis is crucial to understanding the progenitor systems and explosion scenarios, thereby increasing 
our ability to standardize and use SNe~Ia.

\section*{Acknowledgments}
\label{sec:ack}

We would like to thank many colleagues and collaborators for their support.
In particular, many thanks to Eric Hsiao, Mark Phillips, Nidia Morrell, and the Carnegie Supernova Project (CSP) 
collaboration for invaluable discussion and leading us to the light curve results of \citet{milne10} and 
\citet{brown10}.
The CSP collaboration has provided invaluable opportunities for observational experience and insight into
photometry and spectroscopy of SNe Ia in general.
This research is based on observations using Gemini South.
We would also like to thank the speakers at the Gemini Data Workshop 2010, in Tucson, AZ.
The work presented in this paper has been supported in part by NSF awards AST-0708855 and AST-1008962 
(PI: Peter Hoeflich) and AST-1009464 (PI: Chris Gerardy).
We would also like to express our thanks to Peter van Hoof for creation of {\it The Atomic Line List} (V2.05B18) at 
{\tt http://www.pa.uky.edu/ peter/newpage/} and the NIST ADS Team for creation of {\it The NIST Atomic Spectra 
Database} (V5.2) at {\tt http://physics.nist.gov/asd}.

\facility{Gemini:Gillett(GNIRS)}

\end{document}